\documentclass[twocolumn]{aastex63}

\usepackage{amsmath}
\usepackage{graphicx}
\usepackage{xcolor}

\shorttitle{Pulsar Double-lensing Sheds Light on the Origin of Extreme Scattering Events}
\shortauthors{Zhu, H., ET AL.}
\graphicspath{{./}{figures/}}
\begin{document}

%\title{Long-term measurements of the PSR\,B1133+16 scintillation arcs}
\title{Pulsar Double-lensing Sheds Light on the Origin of Extreme Scattering Events}

\correspondingauthor{Hengrui Zhu}
\email{hz0693@princeton.edu}

\author[0000-0001-9027-4184]{Hengrui Zhu}
\affiliation{Department of Physics \& Astronomy, Oberlin College, Oberlin, OH 44074}
\affiliation{Department of Physics, Princeton University, Jadwin Hall Washington Road, NJ 08544, USA}

\author[0000-0001-7888-3470]{Daniel Baker}
\affiliation{Canadian Institute for Theoretical Astrophysics, University of Toronto, 60 Saint George Street, Toronto, ON M5S 3H8, Canada}

\author[0000-0003-2155-9578]{Ue-Li Pen}
\affiliation{Canadian Institute for Theoretical Astrophysics, University of Toronto, 60 Saint George Street, Toronto, ON M5S 3H8, Canada}
\affiliation{Institute of Astronomy and Astrophysics, Academia Sinica, Astronomy-Mathematics Building, No. 1, Sec. 4, Roosevelt Road, Taipei 10617, Taiwan}
\affiliation{Canadian Institute for Advanced Research, 180 Dundas St West, Toronto, ON M5G 1Z8, Canada}
\affiliation{Dunlap Institute for Astronomy and Astrophysics, University of Toronto, 50 St George Street, Toronto, ON M5S 3H4, Canada}
\affiliation{Perimeter Institute of Theoretical Physics, 31 Caroline Street North, Waterloo, ON N2L 2Y5, Canada}

\author[0000-0002-1797-3277]{Dan R. Stinebring}
\affiliation{Department of Physics \& Astronomy, Oberlin College, Oberlin, OH 44074}

\author[0000-0002-5830-8505]{Marten H. van Kerkwijk}
\affiliation{Department of Astronomy and Astrophysics, University of Toronto, 50 St. George Street, Toronto, ON M5S 3H4, Canada}




\begin{abstract}
In extreme scattering events, the brightness of a compact radio source drops significantly, as light is refracted out of the line of sight by foreground plasma lenses. Despite recent efforts, the nature of these lenses has remained a puzzle, because any roughly round lens would be so highly overpressurized relative to the interstellar medium that it could only exist for about a year.
This, combined with a lack of constraints on distances and velocities, has led to a plethora of theoretical models.
We present observations of a dramatic double-lensing event in pulsar PSR~B0834+06 and use a novel phase-retrieval technique to show that the data can be reproduced remarkably well with a two-screen model: one screen with many small lenses and another with a single, strong one.
We further show that the latter lens is so strong that it would inevitably cause extreme scattering events.
Our observations show that the lens moves slowly and is highly elongated on the sky.
If similarly elongated along the line of sight, as would arise naturally from a sheet of plasma viewed nearly edge-on, no large over-pressure is required and hence the lens could be long-lived.
Our new technique opens up the possibility of probing interstellar plasma structures in detail, leading to understanding crucial for high-precision pulsar timing and the subsequent detection of gravitational waves.
\end{abstract}



\section{Introduction}\label{sec:intro}
Extreme scattering events (ESEs) — propagation-produced variations in quasar flux density — have been a puzzle since their discovery in 1987 \citep{fdjh87}.
ESEs manifest as frequency-dependent changes in the observed flux of quasars, usually a sharp spike followed by a dip, for a period of several weeks to months.

It is now widely agreed that ESEs cannot be explained by intrinsic variations of the source \citep{fdj+94}.
Instead, refraction effects from a dense plasma structure in the ISM, hereafter referred to as a plasma lens, with a length scale of a few astronomical units can explain both the observed flux curve as well as the duration of such events \citep{cfl98}.

Yet one difficulty remains: the required electron density and temperature for a roughly rounded plasma lens implies an over-pressure compared to the diffusive ISM by a factor of $10^3$ \citep{ww98,gs06}.
Such a high pressure indicates that the plasma lens would evaporate on the time scale of a year.
This, combined with a lack of constraints on distances and velocities, has led to a plethora of theoretical models \citep{dpg18}.

%Extreme scattering events are propagation-produced variations in quasar flux density due to inhomogeneity in the interstellar plasma.
%They have been a puzzle since their discovery in 1987 \citep{fdjh87}.

%Recent works, e.g. \citet{bst+16}, highlights their extreme properties and ubiquity.
%Despite recent efforts, the nature of these lenses has remained a puzzle, because any roughly round lens would be so highly overpressurized relative to the interstellar medium that it could only exist for about a year.

It was realized early on that pulsars might be powerful probes of these lenses: pulsars scan the sky quickly and, because of their compact sizes, scintillate due to multi-path scattering in the interstellar medium, yielding many new observables \citep{grb+87,cbl+93,cw86,rlg97}.
In this paper, we present a multi-epoch observation of a double-lensing event in pulsar PSR~B0834+06.
We then use a novel phase-retrieval technique to show that the data can be reproduced remarkably well with a two-screen scattering model: one screen with many small lenses and another with a single, strong one \citep{lpm+16}.
Then, we measure the magnification, size, and velocity of the latter lens, and show that it would inevitably cause extreme scattering events if it passed by the line-of-sight to a quasar.
%Our observations show that the lens moves slowly and is highly elongated on the sky.
%If similarly elongated along the line of sight, as would arise naturally from a sheet of plasma viewed nearly edge-on\citep{rbc87,pk12}, no large over-pressure is required and hence the lens could be long-lived.
%Our new technique opens up the possibility of probing interstellar plasma structures in detail, leading to understanding crucial for high-precision pulsar timing and the subsequent detection of gravitational waves\citep{cbl+93}.

% \mhvk{In principle, could remove this summary to reduce word count.}
The paper is organized as the following: in section \ref{sec:obs} we summarize our observations and review the theory of pulsar scintillation. We describe the phase-retrieval technique we adopted in Section \ref{sec:wf}. The double-lensing model and its agreement with the data are presented in Section~\ref{sec:model}. In Section~\ref{sec:interpret}, we extract parameters of the lens and demonstrate that it is capable of causing Extreme Scattering Events. Lastly, we discuss our results and conclude in Section \ref{sec:conclude}.

\begin{figure}
  \centering
  \includegraphics[width=0.5\textwidth]{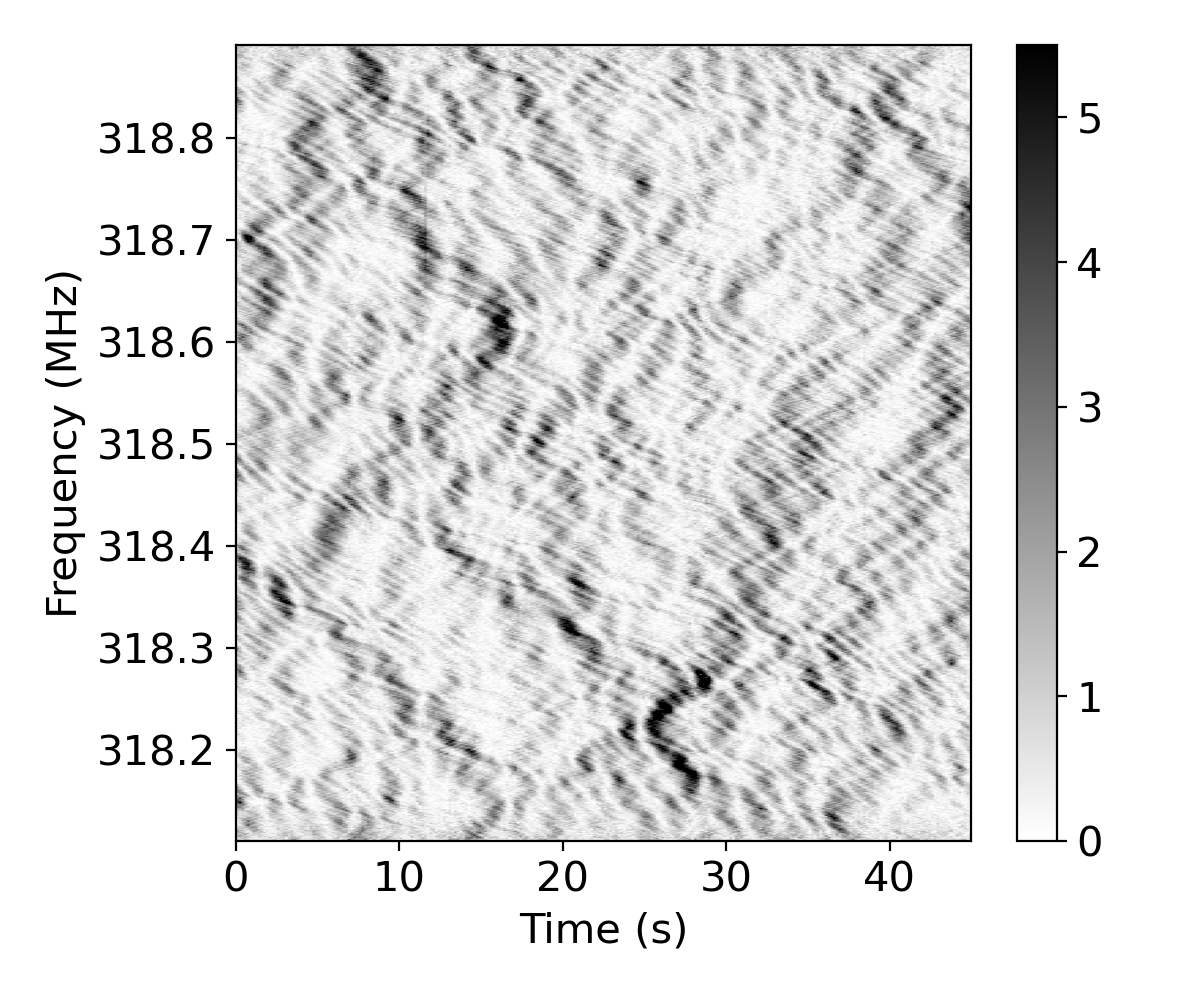}
  \caption{
    Dynamic spectrum for pulsar PSR~B0834+06 for the first observation (MJD 53665).
    Evident is the delicate criss-cross interference pattern resulting from multi-path propagation.
  }
  \label{fig:dspec}
\end{figure}

\begin{figure*}[ht!]
  \includegraphics[width=16.3cm]{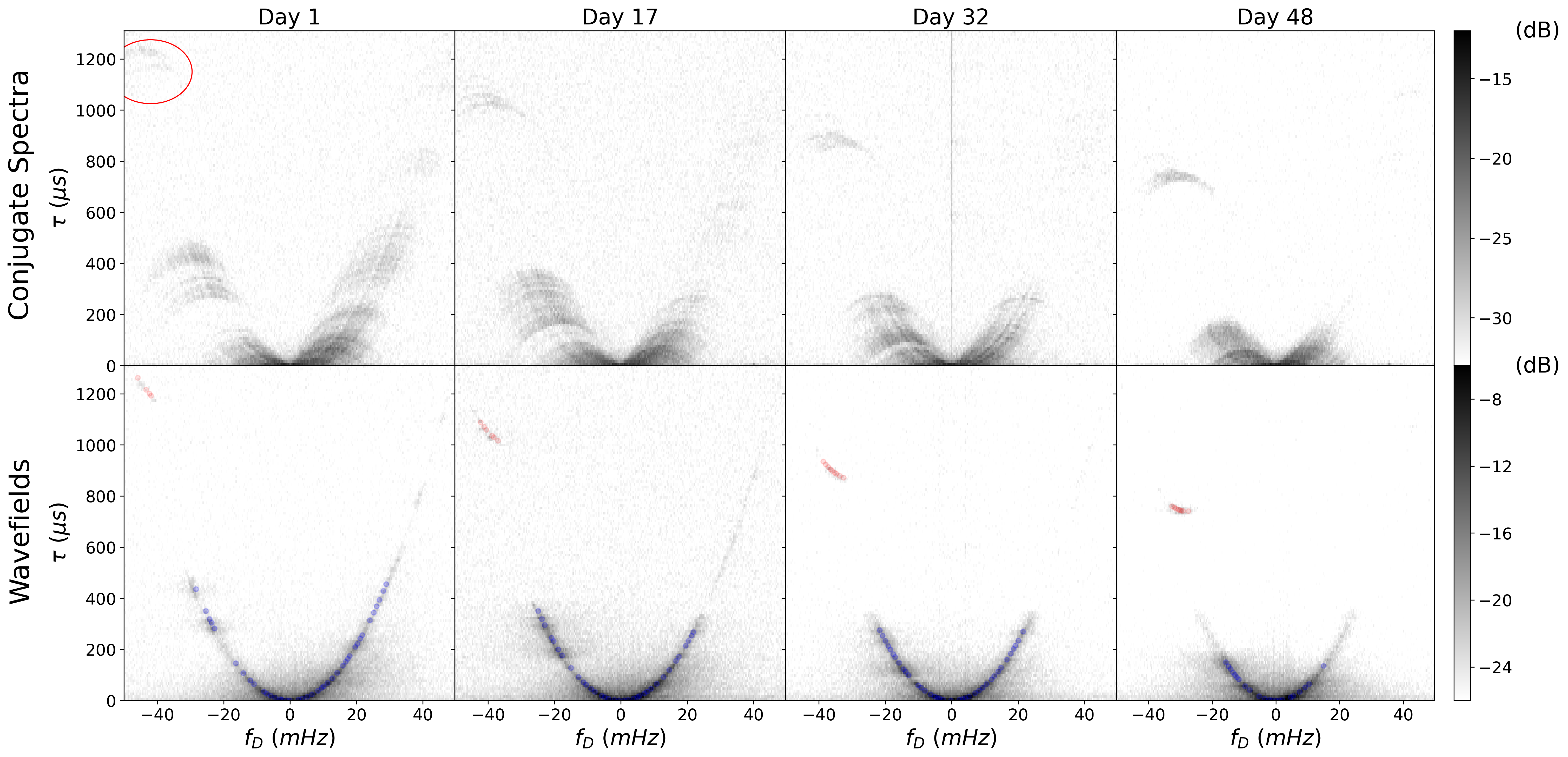}
  \vspace{0.3cm}
  \centering
  \caption{
    Evolution of scintillation structure in PSR~B0834+06 in four equally spaced observations.
    {\em Top:\/} modulus of the conjugate spectra, showing the brightness of the interference between pairs of images on the sky.
    One sees the dominant parabolic scintillation arc, consisting of inverted arclets, as well as, slightly offset from the parabola, the 1~ms feature (circled in red in the left-most panel).  All structures evolve to the right as the pulsar moves on the sky.
    {\em Bottom:\/} modulus of the corresponding wavefields, representing the magnifications from individual paths through the interstellar medium.
    Overlaid are semi-transparent red and blue dots, which correspond to the rays in the lensing model presented in Fig.~\ref{fig:model}, with the blue ones representing rays singly refracted by lenses on the main scattering screen, while the red ones correspond to doubly refracted rays, which pass through both the 1~ms lens and a lens on the main scattering screen.
    Notably, the doubly-refracted images can be reproduced correctly at all epochs with a single lens position.
  }
  \label{fig:sec_wf}
\end{figure*}

\section{Observation and pulsar scintillation}\label{sec:obs}

From October to December 2005, we took seven weekly observations of pulsar B0834+06 at 318-319MHz with the 305-m William E. Gordon Telescope at the Arecibo Observatory.
At each session, we took 45 minutes of data using the 327 MHz receiver with a bandwidth of 0.78 MHz centered at 318.5 MHz.
We created power spectra with 2048 frequency channels, summing the two circular polarizations.
The spectra were then integrated according to the pulsar rotational phase with an integration time of 10 seconds (i.e., averaging over roughly 8 pulses), and the resulting power as a function of frequency, time (10-second chunks), and pulsar rotational phase written out.

We then created dynamic spectra -- power as a function of frequency and time -- by subtracting the background, off-pulse spectrum from the integrated on-pulse signal.
We further divided each time bin of the dynamic spectra by its mean over frequency to mitigate pulse-to-pulse variability of the pulsar.
The spectra show the rich scintillation structures characteristic of scattering in the interstellar medium (Fig.~\ref{fig:dspec}).

To highlight this structure we created conjugate spectra, the Fourier transforms of dynamic spectra, which are functions of Doppler frequency and differential delay, the Fourier conjugates of time and frequency, respectively \citep{crsc06}.
We show the modulus of four of our conjugate spectra in the top row of Figure~\ref{fig:sec_wf}.
The power in each is concentrated in a broad parabola, called a scintillation arc, which consists of upside-down parabolas called inverted arclets.

At a delay of about 1 ms, there is an island of inverted arclets in the conjugate spectrum that migrates consistently down and to the right during the 7 weeks.
This ``1~ms feature'' was first reported by \cite{bmg+10} in a single-epoch VLBI observation made in the middle of our set of observations, and follow-up analysis by \cite{lpm+16} suggested it might arise from double lensing of the pulsar.
As will become clear below, we find that the conjugate spectra can be modelled in detail using a double-lensing interpretation, and that the feature arises from a surprisingly strong lens.

The main scintillation arc can be understood from considering pairs of scattered images of the pulsar.
% see, e.g. \citet{smc+01,crsc06}.
In terms of their (complex) magnifications $\mu_{j,k}$ and angular offsets $\boldsymbol{\theta}_{j,k}$ from the line of sight, the relative Doppler frequency $f_D$, geometric delay $\tau$, and brightness $I$ of a given pair $j,k$ are given by,
\begin{align}
    \tau &= f_{\nu} = \frac{d_{\rm eff}}{2c} (|\boldsymbol{\theta}_j|^2-|\boldsymbol{\theta}_k|^2),\label{eq:tau_def}\\
    f_{D} &= f_{t} = \frac{1}{\lambda}(\boldsymbol{\theta}_j-\boldsymbol{\theta}_k) \cdot \mathbf{v}_{\rm eff}~\label{eq:fd_def},\\
    I &= |C(\tau, f_D)| = |\mu_j \mu_k^*|,\label{eq:ijk_def}
 \end{align}\noindent
where $d_{\rm eff}$ and $\mathbf{v}_{\text{eff}}$ are the effective distance and velocity of the pulsar-screen-Earth system, $C$ is the conjugate spectrum, $c$ is the speed of light, and $\lambda$ is the observation wavelength (see Appendix~\ref{app:theory} for details).

Generally, the magnifications are largest near the line of sight, and thus the brightest signals arise when one member of the pair has $|\boldsymbol{\theta}|\simeq0$.
Considering this, one infers $\tau\propto f_D^2$, thus reproducing the main parabola, as long as $d_{\rm eff}$ and $\mathbf{v}_{\rm eff}$ are roughly constant, which would happen in a thin-screen scattering geometry with scattering localized along the line of sight \citep{smc+01,crsc06}.

The inverted arclets arise from mutual interference between the scattered images, and their sharpness is a signature of highly anisotropic scattering, where the scattered images of the pulsar lie along a (nearly) straight line \citep{wmsz04}.
Like in previous observations, we find that the arclets move along the scintillation arc at a constant speed for long periods of time \citep{hsa+05,msm+20}.
This implies that the scattered images giving rise to the arclets must arise from a large group of parallel and elongated structures in the scattering screen, e.g., turbulence elongated along a given direction \citep{gs95}, waves on a plasma sheet seen in projection as folds \citep{rbc87,pl14}, or magnetic noodles of plasma stabilized by reconnection \citep{gwi19}.

%As can be seen in the example shown in Fig.\ref{fig:dspec}, the resulting spectra show intricate scintillation patterns.
%The scintillation is fully modulated, with the standard deviation of the dynamic spectrum equal to its mean.
%And stationary-phase approximation, resulting in ray optics, has been productive in the regime \citep{wmsz04m,gbr+98m}.
%This implies strong scintillation, where the scattering can be approximated by ray optics.

%We created conjugate spectra using Fourier transforms, which, given the frequency and time resolution of the dynamic spectra, have Nyquist rates of
%$1.3{\rm\,ms}$ in delay and $50{\rm\,mHz}$ in Doppler frequency, sufficient to capture all scintillation structure.
%The resulting conjugate spectra, presented in Fig.~\ref{fig:sec_wf}, show rich arclet structures, which imply a highly anisotropic scattering scenario \citep{wmsz04m,crsc06m}.

%\input{Obs_table.tex}

\section{Phase Retrieval.}\label{sec:wf}

Scattering by the interstellar medium can be well-described as a linear filter.
Hence, the observed signal is the convolution between the impulse response function of the interstellar medium and the intrinsic pulsar signal \citep{wksv08,ws05m,wdv13m,pmdb14m}.
However, because pulsar emission is like amplitude-modulated noise, for slow pulsars whose pulse width is longer than the scattering time, the observed signal contains no useful phase information.
Instead, via the dynamic spectrum one only has a measurement of the squared modulus of the impulse response function.

In general, retrieving the phases from just the amplitudes is an ill-posed problem.
However, when the scattering is highly anisotropic, phase retrieval becomes possible.
The method we use is described in detail in \citet{bvm+21}, but, briefly, it relies on two realizations.
First, for highly anisotropic scattering, the vector offsets $\boldsymbol{\theta}_{j,k}$ in Equations~\ref{eq:tau_def} and~\ref{eq:fd_def} become effectively one-dimensional and for any given $\eta$ in $\tau=\eta f_D^2$ it becomes possible to remap the conjugate spectrum $C(f_D, \tau)$ to $C(\theta_j, \theta_k)$.
For the correct quadratic constant of proportionality $\eta$, which is also the curvature of the scintillation arc, one then finds that the main arc and the arclets are aligned with the cardinal directions in $\theta-\theta$ space \citep{swm+20m}.
Second, if aligned, $C(\theta_j,\theta_k)$ can be factorized using eigenvector decomposition, and the largest eigenvector will be an estimate of the impulse response function \citep{bvm+21}.

In our determinations of the wavefields from each of our dynamic spectra, we follow the procedure of \citet{bvm+21} in detail.
In particular, we reduce the frequency resolution of the dynamic spectra in order to avoid including information from the 1~ms feature (for which the assumption of a single one-dimensional screen does not hold).
We then map the resulting impulse response function estimates back to estimates of the dynamic wavefield, interpolating to the original resolution, and replace the estimated amplitudes with those given by the observed dynamic spectrum.
Finally, Fourier transforms of the estimated dynamic wavefields yield the wavefields in the frequency domain, presented in the second row of Fig~\ref{fig:sec_wf}.
%Recently, it was found that in highly anisotropic scattering, it becomes possible to factorize the conjugate spectrum \citep{swm+20,bvm+21}, creating a ``wavefield,'' which contains the individual magnifications $\mu$ rather than the interference between all pairs of scattered images \citep{wksv08}.
%We applied this technique to our observations and found it worked very well (see Appendix \ref{app:wf}).
%The resulting wavefields are shown in the bottom row of Fig.~\ref{fig:sec_wf}:

We see that in Fig~\ref{fig:sec_wf}, all arclets are reduced to single points in the wavefields, with most lying along the parabola from the main screen, and a few in the 1~ms feature. The power in the 1~ms feature concentrates along a blizzard shape that resembles a portion of a parabola, offset from the origin. We will show that such structure is exactly as predicted by the double-lensing geometry.
%Such wavefield representation reduces redundant information in the conjugate spectra and allows for a precise test of the lensing geometry.

\begin{figure*}[ht!]%
  \includegraphics[width=16.3cm]{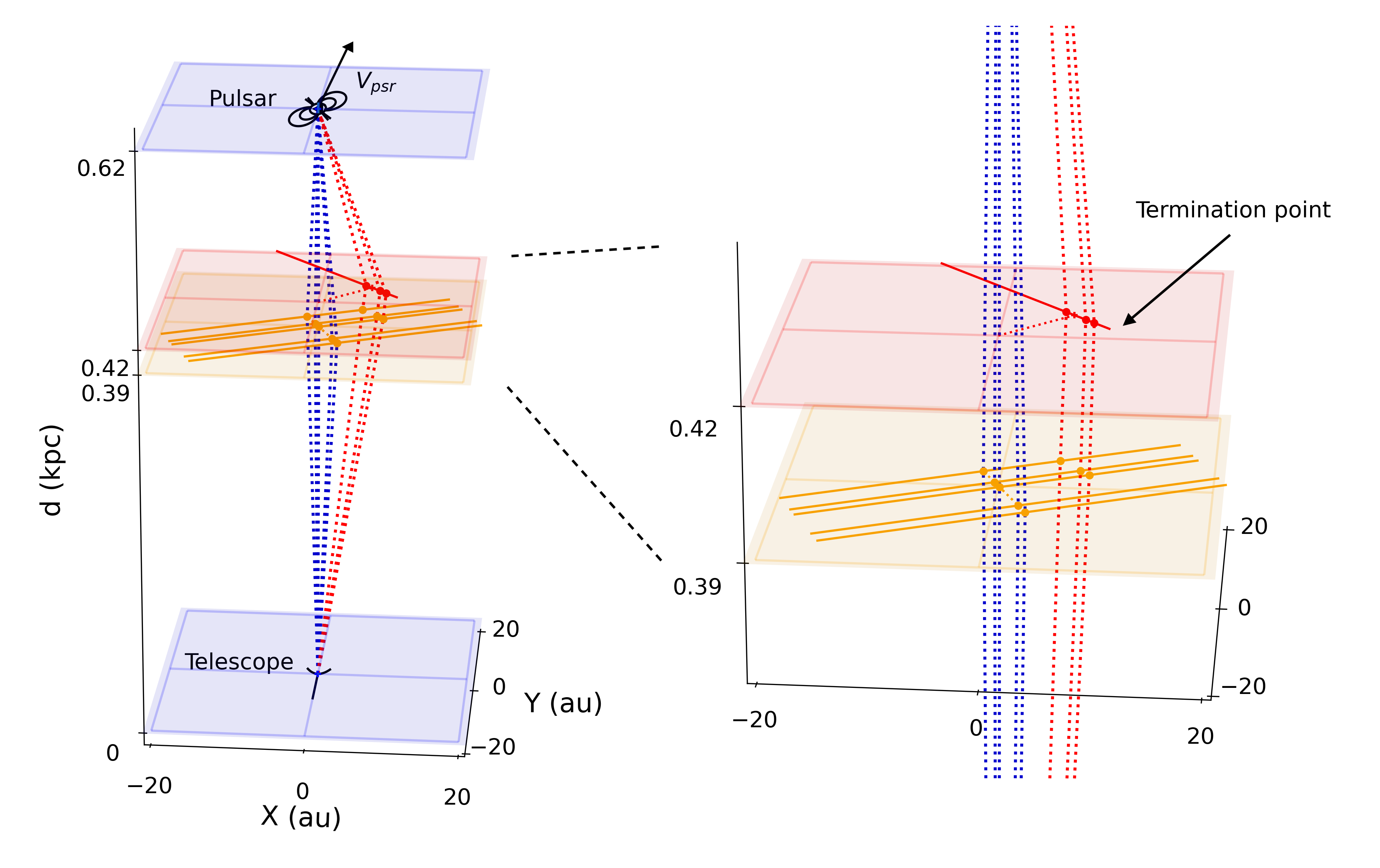}
  \centering
  \caption{Double lensing geometry.
    Shown are the pulsar and observer planes (blue) as well as those of the main scattering screen (orange) and the 1~ms feature (red), with greatly enlarged scale in  X and Y (along right ascension and declination, respectively).
    The highly anisotropic lenses are represented by line segments; they can bend rays only perpendicular to themselves.
    For clarity, only a few of the main-screen lenses are shown.
    Rays between the pulsar and Earth are shown by dashed lines; those bent only by the main screen (blue) give rise to the main scintillation arc, while those doubly refracted (red) lead to the 1~ms feature.
    Note that the 1~ms lens terminates (see expanded view and text).} \label{fig:model}
\end{figure*}

\section{Double-lensing model}\label{sec:model}
With the wavefield, it becomes possible to test the double lensing model directly.
We construct a simple but detailed model, illustrated in Fig.~\ref{fig:model}, in which we represent individual lenses as linear structures that can bend light only perpendicular to their extent.
We assume two lensing planes, with velocities, distances, and orientations of the linear lenses taken from \citet{lpm+16} (see Table~\ref{tab:1}).
Next, for each epoch we use the local maxima along the scintillation arc in the wavefields to determine the location of the linear lenses on the main lensing screen.
Given those locations, the doubly refracted rays can then be solved and the corresponding relative Doppler frequency and delay calculated (see Appendix~\ref{app:double_lens}).

In Fig.~\ref{fig:sec_wf}, we show with faint blue dots the points on the main arc that we use to define linear lenses on the main scattering screen, and with red dots the corresponding double-lensed rays inferred from our model (blue and red rays in the model in Fig.~\ref{fig:model}, resp.).
As can be seen in Fig.~\ref{fig:sec_wf}, the model perfectly reproduces the observed 1~ms feature, including its evolution over a period of 50 days, indicating the 1~ms feature indeed arises from double refraction by highly anisotropic lenses.

As noted above, the images making up the 1~ms feature form part of a parabola offset from the origin.
That the parabola is incomplete implies that not all lenses in the main screen participate in the double refraction, confirming the conclusion of \cite{lpm+16} that the lens that causes the 1~ms feature terminates.
The required geometry is illustrated in the right panel of Fig.~\ref{fig:model}: the two orange lenses at the bottom do not contribute double-lensed rays (red, dashed lines) because of the termination of the 1~ms (red) lens.
As time progresses, not only the overall delay decreases while the pulsar moves towards the 1~ms lens, but also more and more of the bottom of the parabola appears.
This entire evolution of the 1~ms feature is captured by our simple double-lensing model, as shown in the bottom row of Fig.~\ref{fig:sec_wf}.

\begin{table}
\begin{center}
\begin{tabular}{ll}
\hline
\hline
\textbf{Parameter} & \textbf{Value} \\
\hline
%$\parallel$ Scattering axis  & $-25.2 \pm 0.5$ deg east of north \\
%$\perp$ Scattering axis & $-115.2 \pm 0.5$ deg east of north \\
%\hline
$d_{\rm psr}$\dotfill & $620 \pm 60$ pc \\
$\mu_\alpha$\dotfill & $2.16 \pm 0.19$ mas/yr\\
$\mu_\delta$\dotfill & $51.64 \pm 0.13$ mas/yr\\[.8ex]
$d_1$\dotfill  &  $389 \pm 5$ pc \\
$d_2$\dotfill  &  $415 \pm 11$ pc \\
$v_{\rm 1 \parallel}$\dotfill & $-23 \pm 1$ {\rm km/s} \\
$v_{\rm 2 \parallel}$\dotfill & $-3 \pm 3$ {\rm km/s} \\
$\alpha_1$\dotfill & $154.8 \pm 1$ deg\\
$\alpha_2$\dotfill & $136.1 \pm 1$ deg\\
\hline
\end{tabular}
\end{center}
\caption{
  Distances, velocities, and orientations from \citet{lpm+16}.
  The pulsar distance ($d_{\rm psr}$) and angular velocity towards East and North ($\mu_\alpha$ and $\mu_\delta$) are measured from VLBI observations.
  The distances to the main scattering screen ($d_1$) and that of the 1~ms lens ($d_2$) are calculated assuming the pulsar velocity and distance, and their uncertainties reflect only the uncertainty in the relative distance and velocity.
  For the velocities, we can only constrain the component parallel to the images (i.e., along the direction normal to the linear lenses).
  The position angles $\alpha_1$ and $\alpha_2$ are between the lines of images and North (through East).
  The central value of $\alpha_2$ was adjusted slightly (within the uncertainty) from that given in \citet{lpm+16} to best reproduce the wavefields presented in Fig.~\ref{fig:sec_wf}.
}
\label{tab:1}
\end{table}

\begin{figure}[!t]
  \centering
      \includegraphics[width=\hsize]{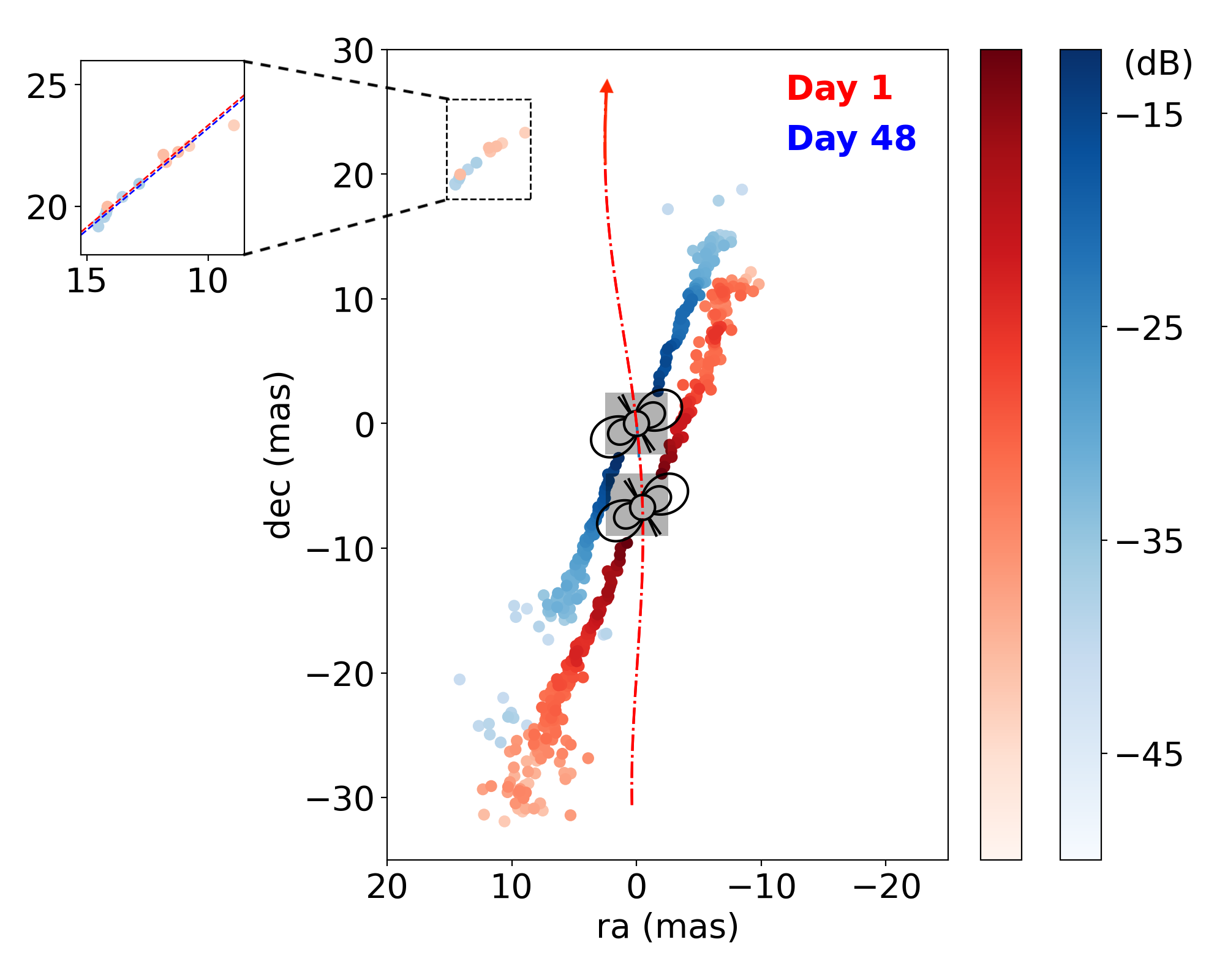} %}\hfill
  \caption{
    Holographic images of the scattered light from the pulsar in the first (red) and last (blue) epoch.
    The trajectory of the pulsar, reflecting its North-ward proper motion and parallax, is shown by the red-dashed curve.
    The images are localized assuming all scattering occurs at the main scattering screen.
    Most lie along a straight line, but those associated with the 1~ms feature are offset.
    As the latter are doubly-refracted, they move with the 1~ms lens rather than with the pulsar.
    The small (if any) offset between the two epochs implies a slow lens velocity (inferred quantitatively from the linear fits shown in the enlarged view).
    Note that the holographic technique has very poor angular resolution close to the pulsar, and hence we do not show results in the grey box with pulsar sketch.
  }
  \label{fig:1ms_motion}
  %\vspace*{0.5cm}}
\end{figure}

\section{Properties of the 1 ms Lens}
\subsection{Velocity and Aspect Ratio}
From the VLBI observations of \cite{bmg+10}, \cite{lpm+16} inferred a low velocity of both lenses.
We can confirm this from our wavefield spectra, as those spectra are essentially holographic images of the pulsar in delay-Doppler-shift space.
This means that for a single screen, given knowledge of the velocities and distances of the pulsar, screen, and Earth, the wavefield can be mapped to the lensed image of the pulsar on the sky, with only a reflection ambiguity around the direction of motion.
One generally cannot map a second screen using the same procedure, but in our case the two scattering planes are known to be much closer to each other than they are to Earth or the pulsar (see Table~\ref{tab:1}), and hence the holographic mapping still yields a reasonable approximation also for the 1~ms feature.

%Furthermore, the reflection degeneracy was resolved by the previous VLBI observation \citep{bmg+10}.

%The properties of the 1~ms feature can be extracted by mapping the wavefields to images on the sky.
%The wavefield contains information on the brightness of all scattered images relative to the line-of-sight image, and for a single screen
The construction of the holographic images makes use of the fact that for the wavefield one has (see App.~\ref{app:theory}),
\begin{eqnarray}
  \tau&=&\frac{d_{\rm eff}}{2c}|\boldsymbol\theta|^2,\\
  f_{D}&=&\frac{\boldsymbol\theta\cdot\boldsymbol{v}_{\rm eff}}{\lambda}.
\end{eqnarray}
Thus, geometrically, the Doppler frequency $f_{D}$ constrains the scattered image to a line on the sky perpendicular to the effective velocity, whereas the differential delay $\tau$ constrains it to a circle centered on the line-of-sight image.
Since the line intersects with the circle twice, at two points symmetric around the direction of the effective velocity, the mapping has a two-fold ambiguity.
In our dataset, this ambiguity is resolved by the previous VLBI observation of \cite{bmg+10}\footnote{A sign error in the analysis of \cite{bmg+10} caused their VLBI images to be flipped along both axis.
  As a result, the 1~ms feature was mapped to the South of the pulsar, which is inconsistent with its delay decreasing with time.
  This sign error is important here, but does not influence the discussion in \cite{bmg+10}}.

To produce the scattered images of the pulsar, we first rebinned our wavefield power spectra by a factor of 4 in delay to increase the signal-to-noise ratio.
Then, for each delay, we subtracted the noise floor and calculated average fluxes and Doppler frequencies inside masked regions around the main parabola (on both positive and negative sides) and around the 1~ms feature, and then converted delay and Doppler frequency to angles along and perpendicular to the direction of $v_{\rm eff}$ using the above equations.

We present the resulting approximate pulsar images of the first and last epochs in Fig.~\ref{fig:1ms_motion}.
The scattered images mostly lie along a straight line, but the 1~ms feature is mapped onto a line segment in a different direction.
While both structures are linear, the mechanisms for producing them differ.
The main linear group of images arises from singly scattered rays by a parallel set of linear lenses.
Each linear lens creates an image at a location closest to the line of sight, so the images move with the pulsar, and the lenses are extended perpendicular to this linear group of images (see Fig.~\ref{fig:model}).
On the other hand, the 1~ms feature is created by double lensing, in which rays are partially bent by the 1~ms lens and then further bent towards the observer by lenses on the main screen.
As seen from Earth, these images can move along the 1~ms lens as the pulsar moves, but not perpendicular to it.
Without the main scattering screen, the 1~ms lens would only be capable of producing a single scattered image (an image that we do not see because the 1~ms lens terminates, although it should have appeared shortly after our campaign, when the pulsar crossed the termination point).

Since the multiple images we see from the double lensing trace out part of the 1~ms lens, they reveal directly that it has a high aspect ratio.
The lack of movement between first and last epoch also shows that the lens velocity is small.
To quantity this, we fitted straight lines to the images for both epochs.
From the perpendicular shift between the lines, we infer an upper bound of $3{\rm\,km \,s^{-1}}$ on the velocity component parallel to its normal, consistent with inferences of \cite{lpm+16} from the VLBI results (see Table~\ref{tab:1}).

\subsection{Width and Magnification of the Lens}\label{sec:width}
Another important parameter for the 1~ms lens is its width, which we can estimate from the magnification~$\mu$.
In coordinates centred on the lens, given a true angular position $\beta$ of a source and an apparent position $\theta$ of its refracted image, by conservation of surface brightness, the magnification is given by \citep{sp18,cfl98},
\begin{equation}
    \mu = \frac{\mathrm{d}\theta}{\mathrm{d}\beta}.
\end{equation}
For a lens far away from the source and $\mu$ not too large, one can approximate $\mu\sim\theta/\beta$, estimate the width $\omega$ of the lens as $\omega\sim\theta\sim\mu\beta$, and use that $\theta\ll\beta$, so that the observed offset $\delta\beta = \beta-\theta\sim\beta$ and thus $\omega\sim\mu\delta\beta$.

The magnification of a doubly refracted image equals the product of the magnifications by the two lenses.
To measure the magnification for the 1~ms lens, we measured the fractional flux of the 1~ms feature and a region on the main scintillation arc associated with the lenses on the main screen that participated in the double refraction (see Appendix~\ref{app:flux}).
Averaging the values for the different epochs, we find that the 1~ms lens has $\mu = 0.06\pm0.02$.
Combined with $\delta\beta = 24\pm4\,\mathrm{mas}$, the inferred angular width of the lens is then $\omega=1.5\pm0.5{\rm\,mas}$, corresponding to a physical width  $w=0.6\pm0.2{\rm\,AU}$.

\section{Interpretation \& Discussion}\label{sec:interpret}

We argue that the parameters we measured for the 1~ms lens imply that it would cause extreme scattering events if it passed in front of a quasar.
%Furthermore, we show that the number density of the lenses as inferred from previous observations do reflects the duty cycle of ESEs.
First, for any lens to cause a significant drop in flux, it must deflect a radio source by more than half of its angular width.
The 1~ms lens, given the maximal delay seen in our observations, is able to deflect radiation by at least 83~mas at 318~MHz.
Given that the bending angle scales as the square of the wavelength, it can thus bend light by at least half its width up to 3.4\,GHz, covering the range in frequencies where extreme scattering events are observed.
Second, given its low velocity, the crossing time would be roughly $w / v_\oplus \simeq 40{\rm\,day}$ (where $v_\oplus=30{\rm\,km\,s^{-1}}$ is the orbital velocity of the earth), in agreement with observed extreme scattering events.

The duty cycle of extreme scattering events is about 0.007 \citep{fdjh87}, which means that if lenses like the 1~ms lens are responsible, they cannot be rare: their typical separation would be $\omega/0.007=200{\rm\,mas}$.
For PSR B0834+06, given its proper motion of 50 mas/yr, one would expect it to cross a similar lens roughly every four years.
Of course, many such crossings would be missed, but we note that an event that is (in hindsight) similar to ours occurred in the 1980s \citep{rlg97}.

One may wonder whether some of the more distant lenses on the main screen would also be capable of causing extreme scattering events, since they deflect pulsar radiation by similar angles and some are at least as bright as the 1~ms feature in the holographically reconstructed images shown in Fig.~\ref{fig:1ms_motion}.
However, for the 1~ms lens, the observed brightness of images is much lower than the magnification because rays are refracted twice, while for the lenses on the main screen the brightness is a direct measure of their magnifications.
For those with a large bending angle, we find $\mu\lesssim0.1\%$, and hence infer angular widths below 0.03 mas.
Such widths are small compared with the angular widths of quasars \citep{kmj+18,gkf99}, and hence these lenses cannot cause the significant dimming seen  in extreme scattering events.

If the 1~ms lens is typical of structures causing extreme scattering events, it excludes a number of models.
%In particular, the low velocity eliminates any explosive models that demand the velocity to be orders of magnitude higher \citep{rbc87}, or those that appeal to structures in the Galactic halo \citep{fdjh87}.
In particular, the low velocity eliminates any models that demand the velocity to be orders of magnitude higher, like those that appeal to structures in the Galactic halo \citep{fdjh87}.
Furthermore, the high aspect ratio undermines isotropic models like large clouds of self-gravitating gas \citep{ww98}.

Indeed, the high aspect ratio may help solve the largest conundrum in extreme scattering events, which is that simple estimates, based on spherical symmetry, give a very high electron density, of $\sim\!10^3{\rm\,cm^{-3}}$, which implies an over-pressure by three orders of magnitude compared to the general interstellar medium \citep{fdjh87,cfl98,ww98}.
For the 1~ms lens, the observed largest bending angle $\alpha\simeq83{\rm\,mas}$ implies a gradient of the electron column density \citep{cfl98},
\begin{equation}
  \frac{{\rm d} N_e}{{\rm d} x} = \frac{2\pi \alpha}{\lambda^2r_e} \simeq 900{\rm\;cm^{-2}\,cm^{-1}},
\end{equation}
where $r_e$ is the classical electron radius.
Thus, under spherical symmetry, one would infer an electron density similar to the problematic ones mentioned above \citep{bst+16}.
Given that the lens is elongated on the sky, however, it may well be elongated along the line of sight too, i.e., be sheet-like.
If so, the inferred electron density within the lens decreases by the elongation factor, and the over-pressure problem can be avoided if the lens is elongated by about a factor~$10^3$.

The above supports the idea that lensing could arise from plasma sheets viewed under grazing incidence.
Both over-dense \citep{rbc87} and under-dense \citep{pk12} sheets have been suggested, and those could be distinguished based on how their behaviour scales with frequency \citep{sp18}.
Unfortunately, for this purpose, our bandwidth of 1\,MHz is too limited, but we believe a reanalysis of the VLBI data on this event will likely yield an answer \citep{bpv22}.

A different test of whether extreme scattering events are caused by structures like our 1~ms lens can be made using quasar flux monitoring surveys: if a lens is highly anisotropic, Earth's orbital motion would cause the lens to be traversed multiple times.
For lenses stationary relative to the local standard of rest, we perform a simulation and find that the likelihood for repeats approaches unity in two antipodal regions on the sky (see Appendix~\ref{app:simulation}).
Monitoring in those two regions can thus further clarify the anisotropy of the lenses responsible for extreme scattering events.

\section{Conclusions}\label{sec:conclude}

In conclusion, we report a strong, slowly-moving and highly anisotropic lens from pulsar scintillation observations.
The observation used only 6 hours of telescope time yet revealed detailed and novel plasma structures in the interstellar medium.
Lenses with properties similar to the 1~ms lens cannot be rare in the Galaxy.
Extended pulsar and quasar monitoring with the next generation survey telescopes like CHIME will further ascertain the statistics of these scatterers, while detailed studies with large telescopes like FAST promise to determine their physical properties.
This will have benefits beyond understanding the lensing proper, since refraction by lenses such as these poses a significant problem for the detection of gravitational waves with a pulsar timing array \citep{cbl+93,cks+15}.
Deeper understanding will help develop mitigation strategies, which will be especially important as we move beyond the detection of a stochastic wave background to an era in which individual gravitational wave sources are analyzed \citep{btc+19}.

\section*{Code availability}
The code used for phase retrieval and generating the wavefields is integrated into the {\tt scintools} package developed by Daniel Reardon \citep{rcb+20}, available at \href{https://github.com/danielreardon/scintools}{\color{blue}github.com/danielreardon/scintools (external link)}. The ray-tracing code used for modeling the double lensing geometry is part of the {\tt screens} package developed by one of us \citep{screens:22} at \href{https://github.com/mhvk/screens}{\color{blue}github.com/mhvk/screens (external link)}.

\section*{Acknowledgements} %
  We dedicate this paper to the Arecibo Observatory and its staff.
  We thank W. Brisken for clarifications regarding his previous VLBI results, Tim Sprenger for discussions and verifying our two-screen solution, and the Toronto scintillometry group for general discussions.
  We appreciate support by the NSF (Physics Frontiers Center grant 2020265 to NANOGrav, and grant 2009759 to Oberlin College; D.S. and H.Z.) and by NSERC (M.H.v.K., U.-L.P. and D.B.).
  We received further support from the Ontario Research Fund—Research Excellence Program (ORF-RE), the Natural Sciences and Engineering Research Council of Canada (NSERC) [funding reference number RGPIN-2019-067, CRD 523638-18, 555585-20], the Canadian Institute for Advanced Research (CIFAR), the Canadian Foundation for Innovation (CFI), the Simons Foundation, Thoth Technology Inc, who owns and operates ARO and contributed significantly to the relevant research, the Alexander von Humboldt Foundation, and the Ministry of Science and Technology of Taiwan [MOST grant 110-2112-M-001-071-MY3].

\software{
  astropy \citep{astropy:13, astropy:18, astropy:22},
  numpy \citep{numpy:20},
  matplotlib \citep{matplotlib:07},
  screens \citep{screens:22}}
 
\bibliography{psrrefs}{}
\appendix

\restartappendixnumbering

\section{Theory of Pulsar Scintillation}\label{app:theory}

Observations suggests that a significant amount of pulsar scattering is dominated by thin screens (for a detailed account, see \citet{smc+01}).
We consider a pulsar behind a single scattering screen, with distances $d_{\rm psr}$ and $d_{\rm scr}$ from the observer, respectively.

The geometric delays for rays that pass through the screen are then the same as for the case that the source is at infinity and the screen is at what is known as the ``effective distance'',
\begin{equation}\label{eq:deff}
    d_{\text{eff}}=\frac{d_{\rm psr}d_{\rm scr}}{d_{\rm psr}-d_{\rm scr}}
    = d_{\rm psr}\frac{1-s}{s},
\end{equation}
where $s = 1-d_s/d_p$ is the fractional distance from the pulsar to the screen.
For two paths at angles $\boldsymbol{\theta}_j$ and $\boldsymbol{\theta}_k$ between the observer and a screen at $d_{\rm eff}$, one then recovers the differential geometric delay:
\begin{equation}
  \tau = f_{\nu} = \frac{d_{\rm eff}}{2c} (|\boldsymbol{\theta}_j|^2-|\boldsymbol{\theta}_k|^2).\label{eq:tau_def2}
\end{equation}
Because of the different delays along different paths, the pulsar's radiation interferes with itself and casts a diffractive pattern onto the observer plane.
The velocity of this diffractive pattern with respect to the observer is called the effective velocity, which determines the differential Doppler shifts:
\begin{equation}
  f_{D} = f_{t} = \frac{1}{\lambda}(\boldsymbol{\theta}_j-\boldsymbol{\theta}_k) \cdot \mathbf{v}_{\rm eff},\label{eq:fd_def2}
\end{equation}\noindent
where the effective velocity is given by,
\begin{equation}\label{eq:veff}
    \mathbf{v}_{\rm eff}= -\frac{1-s}{s}\mathbf{v}_{\rm psr} + \frac{1}{s}\mathbf{v}_{\rm scr} - \mathbf{v}_{\oplus},
\end{equation}\noindent
and $\mathbf{v}_{\rm psr}$, $\mathbf{v}_{\rm scr}$, and $\mathbf{v}_{\oplus}$ are the components perpendicular to the line of sight of the pulsar, screen, and Earth velocities, respectively.

Under the stationary phase approximation, we can treat the pulsar as scattered into $N$ images with (complex) magnifications $\mu_j$ and geometric phase delay $\exp\{i[f_{Dj}t+\tau_j\nu]\}$, where $\tau_j=(d_{\rm eff}/2c)|\boldsymbol\theta_j|^2$ and $f_{D_j}=(\boldsymbol\theta_j\cdot \boldsymbol{v}_{\rm eff})/\lambda$ are the geometric delay and Doppler frequency with respect to the line-of-sight image.
% \begin{align}
%     \tau_j &= \frac{d_{\rm eff}}{2c} |\boldsymbol{\theta}_j|^2,\label{eq:tauj_def}\\
%     f_{D_j} &= \frac{1}{\lambda}\boldsymbol{\theta}_j\cdot \mathbf{v}_{\rm eff}~\label{eq:fdj_def}.
% \end{align}
% These images correspond to stationary phase points on the scattering screen.
We normalize the flux so that $\sum_j|\mu_j|^2=1$.
% \begin{equation}
%     \sum_j\mu_j^2 = 1~.
% \end{equation}

The dynamic spectrum, which encodes the interference of the $N$ images, is given by,
\begin{align}
    D(t,\nu) &= \left\vert\sum_j \mu_j\exp\{2\pi i[f_{Dj}t+\tau_j\nu]\}\right\vert^2~\\
    &= \sum_{j,k}\mu_j\mu_k\exp\{2\pi i[(f_{Dj}-f_{Dk})t+(\tau_j-\tau_k)\nu]\}.
\end{align}
Note that the dynamic spectrum defined above is entirely real as the phase is antisymmetric under the exchange of j and k.
The Fourier transform of the dynamic spectrum is the conjugate spectrum; its square modulus, called the secondary spectrum, is given by,
\begin{equation}
    |C(f_D,\tau)|^2 = 2\sum_{j,k}\mu_j^2\mu_k^2 \delta(f_D,f_{D_j}-f_{D_k}) \delta(\tau,\tau_j-\tau_k).\label{eq:ss}
\end{equation}

\section{Double refraction by two linear lenses.}\label{app:double_lens}
Consider two screens with linear features between the telescope and the pulsar, and a cylindrical coordinate system in which $z$ is along the line of sight, with direction $\hat{z}$ pointing towards the pulsar.
For scattering screen $i$ at distance $d_i$, a line, representing a linear lens on screen $i$, can be written as,
\begin{equation}
d_{i}\hat{z} + \vec{r}_{i} + \sigma \hat{u}_{i},
\end{equation}\noindent
where $\vec{r}_{i}$ is a cylindrical radius from the line of sight to
the line (i.e., $\hat{r}_i\cdot\hat{z}=0$), $\hat{u}_i=\hat{z}\times\hat{r}_i$ a
unit vector perpendicular to it in the plane of the screen, and $\sigma$ is the position along the line.

Imagine now a ray going from the observer to some point along a linear lens on the first screen, at distance $d_1$.
Since it will be easiest to work in terms of angles relative to the observer, we use $\rho=r/d$ and $\varsigma=\sigma/d$ to write this trajectory, from $d = 0$ to $d = d_1$, as,
\begin{equation}
d(\hat{z} + \rho_{1}\hat{r}_{1} + \varsigma_{1}\hat{u}_{1}).
\end{equation}
When the ray hits the lens, light can be bent only perpendicular to the lens, by an angle which we will label $\alpha_1$ (with positive $\alpha_1$ implying bending closer to the line of sight \citep{sp18}).
Hence, beyond the screen, for $d > d_1$, its trajectory will be
\begin{equation}
  d(\hat{z} + \rho_{1}\hat{r}_{1} + \varsigma_{1}\hat{u}_{1})
  - (d-d_{1})\alpha_{1}\hat{r}_{1}.
\end{equation}
If the ray then hits a lens on the second screen at a distance $d_2$, it will again be bent, by $\alpha_2$, and then follow, for $d>d_2>d_1$,
\begin{equation}
  d(\hat{z} + \rho_{1}\hat{r}_{1} + \varsigma_{1}\hat{u}_{1})
  - (d-d_{1})\alpha_{1}\hat{r}_{1}
  - (d-d_{2})\alpha_2\hat{r}_{2}.
\end{equation}
In order to specify the full trajectory, we need to make sure that the ray actually intersects the lens on the second screen, and ends at the pulsar, i.e.,
\begin{align}
    d_{2}(\hat{z} + \rho_{1}\hat{r}_{1} + \varsigma_{1}\hat{u}_{1}) - (d_{2}-d_{1})\alpha_{1}\hat{r}_{1}
     &= d_{2}(\hat{z} + \rho_{2}\hat{r}_{2} + \varsigma_{2}\hat{u}_{2}),\\
    d_{p}(\hat{z} + \rho_{1}\hat{r}_{1} + \varsigma_{1}\hat{u}_{1}) - (d_{p}-d_{1})\alpha_{1}\hat{r}_{1} - (d_{p}-d_{2})\alpha_{2}\hat{r}_{2}
     &= d_{p}\hat{z}.
\end{align}
These constraints can be simplified to,
\begin{align}
\varsigma_{1}\hat{u}_{1} - (1-d_{1}/d_{2})\alpha_{1}\hat{r}_{1} - \varsigma_{2}\hat{u}_{2}
 &= \rho_{2}\hat{r}_{2} - \rho_{1}\hat{r}_{1},\label{eq:sim_1}\\
\varsigma_{1}\hat{u}_{1} - (1-d_{1}/d_{p})\alpha_{1}\hat{r}_{1} - (1-d_{2}/d_{p})\alpha_2\hat{r}_{2} &= -\rho_{1}\hat{r}_{1}.\label{eq:sim_2}
\end{align}
Hence, one is left with a pair of two-dimensional vector equations with four scalar unknowns, viz., the bending angles $\alpha_{1,2}$ and angular offsets $\varsigma_{1,2}$ along the two linear lenses.
This set of equations can be extended easily to arbitrary number of screens and a possible non-zero origin.
Since the equations are linear in the unknowns, the set can be solved using matrix inversion.
We use the {\tt screens} package \citep{screens:22} for this purpose, which also uses the inverted matrix to calculate time derivatives $\dot\varsigma$ and $\dot\alpha$ given velocities of the observer, screens, and pulsar (and thus $\dot\rho$ in the equations), as well as the implied delay $\tau$ and its time derivative $\dot\tau$ for each ray.

% Knowing the geometry of a doubly refracted ray, we can calculate its delay and Doppler frequency with respect to the line-of-sight image via \citep{spmb19},
% \begin{align}\label{eq:2s_tau}
% \tau_{12} &= \frac{1}{2c} \left[\left(\frac{1}{d_\mathrm{psr}-d_1}  +\frac{1}{d_1-d_2}\right) d_1^2|\boldsymbol{\theta}_1|^2 + \left(1+\frac{d_2}{d_1-d_2}\right)d_2 |\boldsymbol{\theta}_2|^2- \frac{2d_1d_2}{d_1-d_2}\boldsymbol{\theta}_1\cdot\boldsymbol{\theta}_2\right],\\
% f_{D,12} &= \frac{1}{\lambda}\bigg[
% (\mathbf{v}_{2r}-\mathbf{v}_\oplus)\cdot\boldsymbol{\theta}_2 +\frac{1}{d_1-d_2}(d_1\boldsymbol{\theta}_1-d_2\boldsymbol{\theta}_2)\cdot(\mathbf{v}_{1r}-\mathbf{v}_{2r}) + \frac{d_1}{d_p-d_1}\boldsymbol{\theta}_1\cdot(\mathbf{v}_{1r}-\mathbf{v}_\mathrm{psr})\bigg]
% \end{align}\noindent

% where $\boldsymbol{\theta}_{1,2}$ are the angular coordinates for the intersections of the ray with the two screens, and $\mathbf{v}_{1r,2r}$ the velocities of the intersections, which has complicated dependence on the geometry of the ray as well as velocities in the system.
% Numerical values for $\tau$ and $f_D$ used in this paper are calculated using the {\tt screens} package.

We note that the same scattering geometry is obtained by considering the geometric limit of the two-screen scattering model in \citet{smw+22}.
It is different, however, from the geometry considered by \cite{spmb19}, as those authors assumed that the scattering points were fixed on their respective screens, while we assume that light can only be bent perpendicular to the linear structures, which implies that the scattering images move along those structures as the relative positions of the observer, screens, and pulsar change.

\section{Flux estimation and magnification of the 1~ms lens.}\label{app:flux}
To find the flux of a scattered image $i$, which equals $\mu_i^2$, we integrate over a region corresponding to the interference of that image with all other images, i.e., the corresponding inverted arclet in the secondary spectrum, as defined in equation \ref{eq:ss}.
Evaluating this integral, one finds that it yields twice the flux of image $i$,
\begin{align}
    &\iint \delta(j,i)2\sum_{j,k}\mu_j^2\mu_k^2\delta(f_D, f_{D_j}-f_{D_k})\delta(\tau,\tau_j-\tau_k) {\rm d}f_D {\rm d}\tau\nonumber\\
    &= 2\mu_{i}^2\sum_k\mu_k^2 = 2\mu_{i}^2.
\end{align}

\begin{figure}
  \centering
  \includegraphics[width=0.6\textwidth]{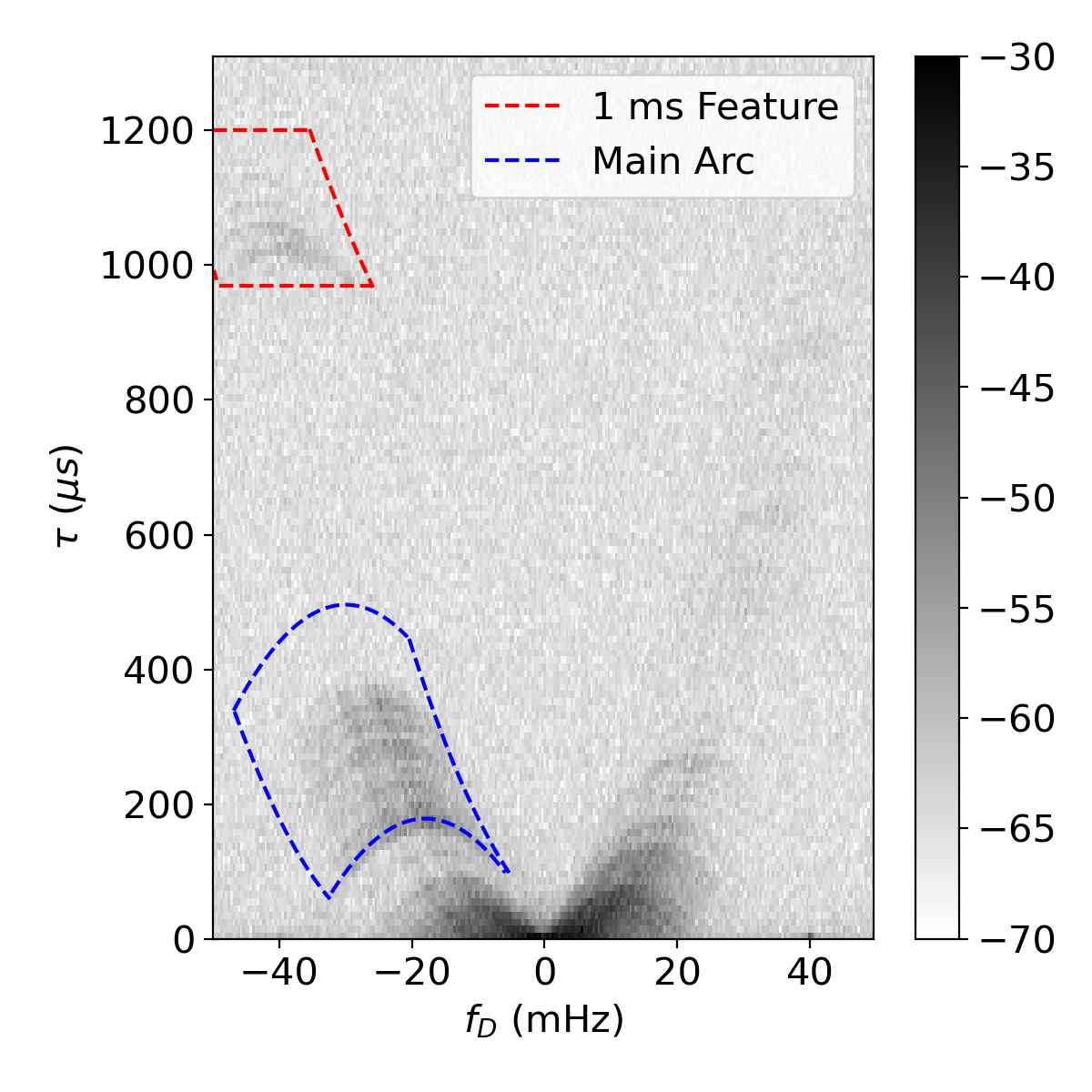}
  \caption{
    Squared modulus of the conjugate spectrum on day 17.
    We estimate the flux for the double refracted radiation and that for the radiation refracted by the main-screen lenses that participated in the double lensing by summing the noise-floor corrected power in the red and blue regions.
  }
  \label{fig:flux_region}
\end{figure}

\begin{figure}
  \centering
  \includegraphics[width=0.6\textwidth]{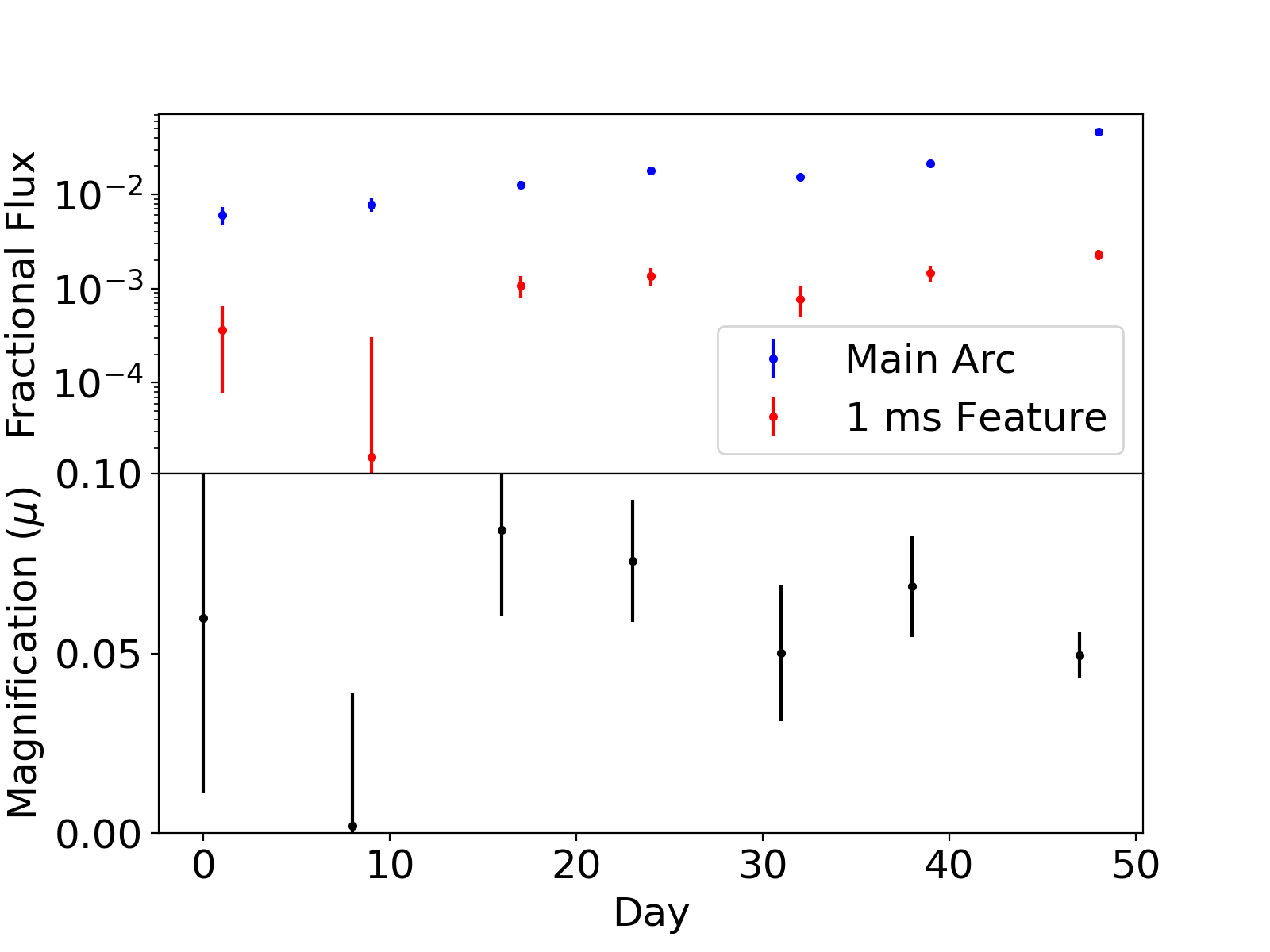}
  \caption{
    Flux and magnification estimates.
    {\em Top:\/} Estimated fractional flux of the 1~ms feature (red) and the corresponding region of the main arc (blue).
    The uncertainties are dominated by the uncertainty with which the noise floor can be estimated; as the non-uniformities are similar in the seven epochs, we use a constant value.
    {\em Bottom:\/} Magnifications for the 1~ms lens estimated from the ratios of the flux for the 1~ms lens and that for the corresponding region on the main arc.
  }
  \label{fig:frac_flux}
\end{figure}

We measured the flux of the 1~ms feature by integrating around it (red region in Fig.~\ref{fig:flux_region}) in each of the 7 secondary spectra (with appropriate noise floors subtracted).
We also measured the flux of the corresponding part of the main arc that participated in the double lensing, as inferred from our double lensing model (blue region in Fig.~\ref{fig:flux_region}).
In Fig.~\ref{fig:frac_flux}, we show the resulting fractional fluxes, as well as their ratio, which we use as an estimate of the magnification of the 1~ms lens.

Generally, both fluxes increase with time, as expected since the images approach the line of sight.
The one exception is the second epoch, in which the 1~ms feature is much fainter.
We do not know the reason for this, but note that on that day the other side of the scintillation arc also appeared much fainter at high delay than it was in the first and third epoch.
Neglecting the second epoch, we find an averaged magnification of $0.06\pm0.02$ for the 1~ms lens, which is what we used in the section~\ref{sec:width} to infer the width of the lens.

Note that one might have expected the magnification of the 1~ms lens to increase with time as its impact parameter got smaller, too.
In our estimate, however, we have ignored that the magnification of the linear lenses on the main screen is not exactly the same for the singly and doubly scattered rays, since these have different bending angles.
We leave a detailed analysis of the magnification as a function of bending angle (and perhaps location along the lens) for future work.

\section{Repetition Likelihoods for Extreme Scattering Events.}\label{app:simulation}
If, as our observations suggest, quasar extreme scattering event are caused by slowly-moving linear plasma lenses, one might expect to see repetitions due to Earth's orbital motion.
The likelihood depends on how the Earth's motion projects on the screen, and thus on the screen orientation and location on the sky.
To estimate the probabilities, we simulated screens over the entire sky, sampling on a Gaussian Legendre grid with 25 points along the declination and 50 points along the right ascension axis.
Then, using the {\tt astropy} package \citep{astropy:18,astropy:13}, we calculated Earth's trajectory relative to the local standard of rest (i.e., orbital motion plus the solar system systemic motion), as projected on the sky for each  angular direction, for a one-year period.
\begin{figure}
  \includegraphics[width=\hsize]{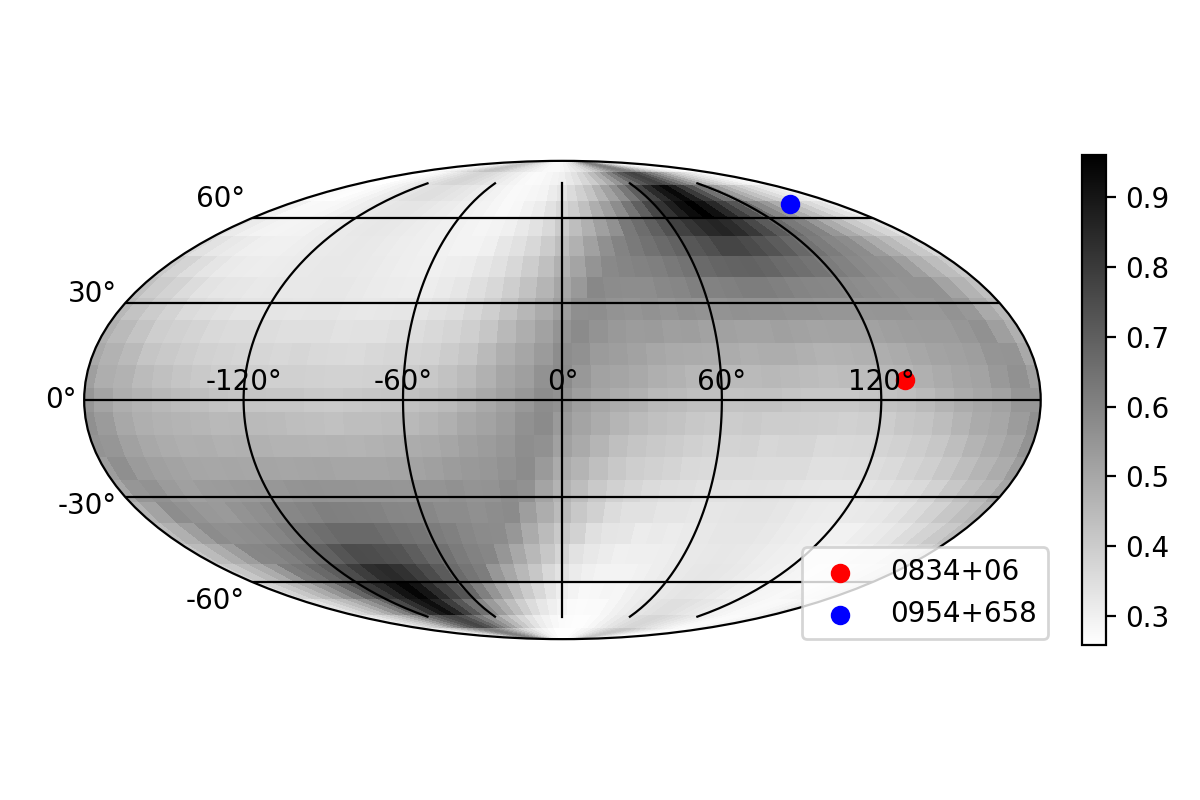}
  \caption{
    Probability for an extreme scattering event to repeat within a year.
    The two antipodal regions with repeat likelihood approaching unity correspond to the directions parallel to the motion of the solar system relative to the local standard of rest, along which Earth's trajectory projected on the sky is a closed ellipse.
    The sinusoidal region with enhanced repitition likelihood arises from the cancellation between Earth's orbital motion and the solar system motion relative to the local standard of rest.
    The locations of the first quasar for which an extreme scattering event was found \citep{fdjh87}, QSO~0954+658, as well as for the pulsar reported here, PSR~B0834+06, are marked with blue and red dots, respectively.}
  \label{fig:repeat_probability}
\end{figure}

We then generated large numbers of randomly oriented lines for each grid point to represent linear lenses, assumed stationary relatively to the local standard of rest.
For each grid point, we counted the number of simulated lenses that intersected more than once with the projected trajectory of Earth, and then took the ratio with the number that intersected at least once as the likelihood for an extreme scattering event to repeat within a year.
We show the result in Fig.~\ref{fig:repeat_probability}.
For the line of sight to the first extreme scattering event, QSO~0954+658, the (interpolated) probability is a modest 0.44, consistent with no repetition having been seen in the three years the source was monitored \citep{fdjh87}, but there are also regions on the sky for which the probability of repeats approaches unity.

\begin{figure}
  \centering
  \includegraphics[width=0.6\textwidth]{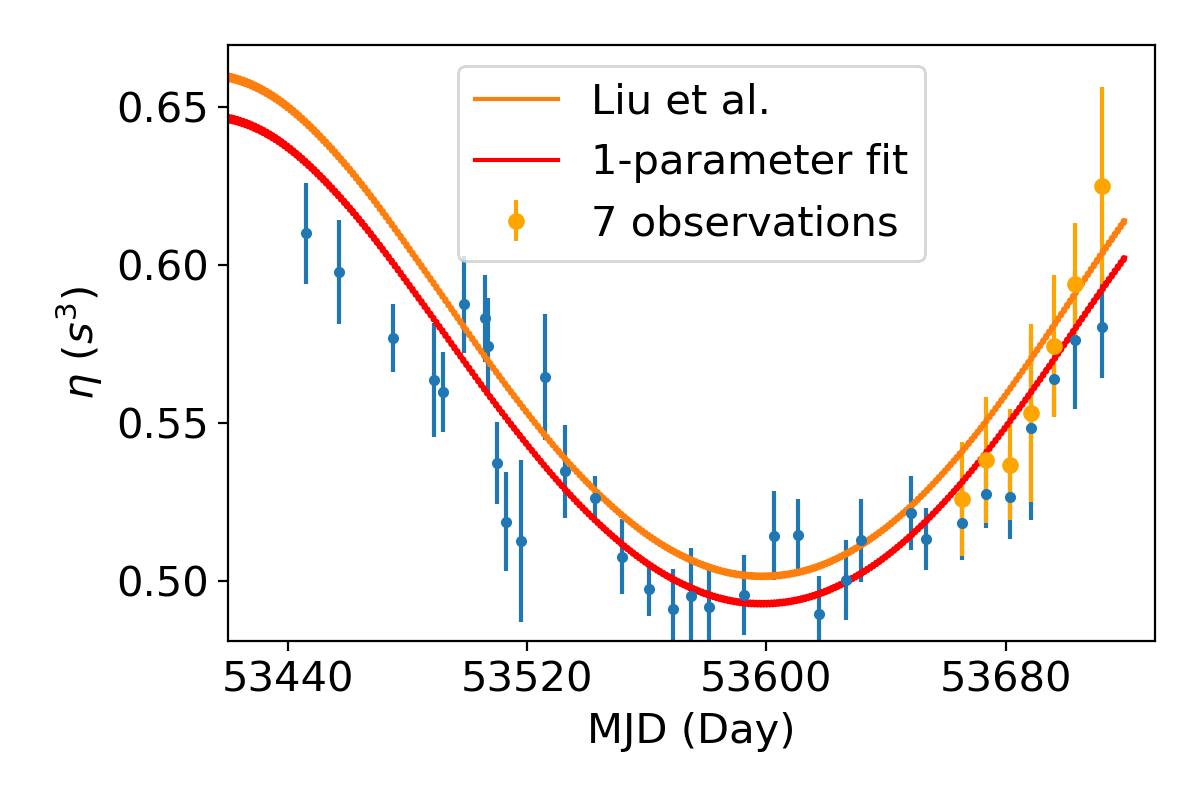}
  \caption{
    Measured quadratic coefficient of the scintillation arc over a year-long observation campaign of PSR~B0834+06.
    The seven observations analyzed here are highlighted in orange.
    The variation predicted using the parameters from \cite{lpm+16} based on previous VLBI work (see Table~1) is slightly offset, but an adjustment of just the screen velocity, within the quoted uncertainty, makes it consistent with our measurements.}
  \label{fig:curv_evo}
\end{figure}

\section{Testing the screen geometry.}\label{app:annuel variation}
The main scintillation arc persists for a long period of time, which offers the opportunity to test whether the velocity and distance to the underlying scattering screen remain consistent with what is inferred from VLBI, using the variation of the quadratic constant of proportionality $\eta$ for the scintillation arc, as induced by the changing Earth orbital motion \citep{mzs+21,spv+21},
\begin{equation}\label{eq:eta2}
  \eta = \frac{d_{\rm eff}\lambda^2}{2c (|\mathbf{v}_{\rm eff}| \cos\alpha)^2},
\end{equation}
where $\alpha$ is the angle between the effective velocity (Eq.~\ref{eq:veff}) and the line defined by the scattered images.

Measured values of $\eta$ for a year-long observing campaign are presented in Fig.~\ref{fig:curv_evo}, with the seven observations analyzed here highlighted.
Overdrawn is the evolution predicted based on Eq.~\ref{eq:eta2} for the pulsar and screen parameters in Table~1 in the main text and the known motion of Earth.
The prediction is qualitatively correct, but somewhat offset from the measurements.
To correct for this offset, we performed a fit in which we only allowed the velocity of the main scattering screen to vary (which is the parameter with the largest fractional uncertainty).
We found a better fit for $v_{1,\parallel}=-24{\rm\,km/s}$, which is within the reported uncertainty.
Thus, overall our results confirm the values inferred from the previous analysis of VLBI data \citep{lpm+16,bmg+10}.

\end{document}